\documentclass[runningheads]{svmult}
\usepackage{graphicx}
\usepackage{subeqnar}
\usepackage{multicol}
\usepackage{rotating}
\usepackage{mathptm}
\usepackage{physprbb}
\def\SNR(#1.#2)#3(#4.#5){{G#1${\kern0.5pt\cdot\kern1.0pt}$#2$#3$#4${\kern0.5pt\cdot\kern1.0pt}$#5}}
\newcommand{\marnote}[1]{\mbox{}\marginpar{\begin{rotate}{-90}
\parbox{20cm}{\small\bf #1}
\end{rotate}}}
\begin{document}
\title*{The Historical Supernovae\marnote{to be published in {\sl ``Supernovae
and Gamma Ray Bursters''}, ed.\ K.W.\ Weiler, Lecture Notes in Physics ({\tt
http://link.springer.de/series/lnpp}),\\Copyright: Springer-Verlag,
Berlin-Heidelberg-New York, 2003}
}
\toctitle{The Historical Supernovae}
\titlerunning{The Historical Supernovae}
\author{David A.\ Green\inst{1}
  \and
 F.\ Richard Stephenson\inst{2}}
\authorrunning{Green \& Stephenson}
\institute{Cavendish Laboratory, University of Cambridge, Madingley Road,
  Cambridge CB3~0HE, United Kingdom
    \and
Department of Physics, University of Durham, Durham DH1~3LE, United Kingdom}

\maketitle

\begin{abstract}
The available historical records of supernovae occurring in our own Galaxy over
the past two thousand years are reviewed. These accounts include the
well-recorded supernovae of AD 1604 (Kepler's SN), 1572 (Tycho's SN), 1181,
1054 (which produced the Crab Nebula) and 1006, together with less certain
events dating back to AD 185. In the case of the supernovae of AD 1604 and 1572
it is European records that provide the most accurate information available,
whereas for earlier supernovae records are principally from East Asian sources.
Also discussed briefly are several spurious supernova candidates, and the
future prospects for studies of historical supernovae.
\end{abstract}

%------------------------------------------------------------------------------%
\section{Introduction}

The investigation of observations of Galactic supernovae is very much an
interdisciplinary exercise, far removed from the usual course of scientific
endeavour. It seems well established that no supernova has been observed in our
own Galaxy since AD 1604. Hence for reports of Galactic supernovae it is
necessary to rely entirely on observations made with the unaided eye. It is
fortunate that early astronomers recorded several of these events -- along with
many other temporary `stars', such as comets and novae. As might be expected
the quality of these early observations is very variable. In the case of the
two most recent Galactic supernovae, in AD 1604 and 1572, European astronomers
measured the positions with remarkable precision -- to within about 1~arcmin --
and the changing brightness was carefully recorded. However, in earlier
centuries the available observations are of lesser quality. Nevertheless,
sightings of three supernovae have been confidently identified from medieval
East Asian records, while three more ancient supernovae have also been
proposed.

Here we review the available historical records of supernovae occurring in our
own Galaxy over the past two thousand years or so. Studies of potential
supernovae from the various historical records will be concentrated on those
new stars which were said to be visible for at least three months.  This
restriction eliminates most novae and also considerably diminishes the
possibility of the object being a comet. More detailed discussions of these
historical supernovae are presented in \cite{Cla77,Ste02}.

In Sect.~\ref{s:good} the available historical records -- from East Asian,
Europe and the Arab Dominions -- of the well-recorded supernovae of AD 1604,
1572, 1181, 1054 and 1006 are discussed. Other probable supernovae occurring
before AD 1000, from Chinese records, are discussed in Sect.~\ref{s:older},
with less convincing possible, and spurious historical supernova discussed in
Sect.~\ref{s:not} (including the suggestion that the supernova that produced
the young SNR Cas~A was seen by Flamsteed in AD 1680). Table~\ref{t:summary}
presents a summary of the well-recorded and probable historical supernovae seen
in our Galaxy, including the sources of historical records. Finally, the future
prospects for studies of historical supernovae are briefly discussed in
Sect.~\ref{s:future}.

\begin{table}
\caption{Summary of the historical supernovae, and the source of their
records}\label{t:summary}
\setlength\tabcolsep{2pt}
\footnotesize
\begin{tabular}{llllccccc}
          & & length of      &                    & \multicolumn{5}{c}{Historical Records}              \\
 date     & & visibility     & remnant            & Chinese & Japanese & Korean & Arabic & European     \\ \hline
 AD1604  & & 12 months      & \SNR(4.5)+(6.8)    &   few   &    --    &  many  &   --   &     many   \\
 AD1572  & & 18 months      & \SNR(120.1)+(2.1)  &   few   &    --    &  two   &   --   &     many   \\
 AD1181  & &  6 months      & 3C58               &   few   &   few    &   --   &   --   &      --    \\
 AD1054  & & 21 months      & Crab Nebula        &  many   &   few    &   --   &   one  &      --    \\
 AD1006  & &  3 years       & SNR327.6$+$14.6    &  many   &  many    &   --   &   few  &     two    \\
 AD393   & &  8 months      & \qquad --          &   one   &    --    &   --   &   --   &      --    \\
 AD386?  & &  3 months      & \qquad --          &   one   &    --    &   --   &   --   &      --    \\
 AD369?  & &  5 months      & \qquad --          &   one   &    --    &   --   &   --   &      --    \\
 AD185   & & 8 or 20 months & \qquad --          &   one   &    --    &   --   &   --   &      --    \\ \hline
\end{tabular}
\end{table}

%------------------------------------------------------------------------------%
\section{Well-defined historical Supernovae}\label{s:good}

\subsection{Kepler's SN of AD 1604}

The new star which appeared in the autumn of AD 1604 was discovered in Europe
on Oct 9, and first noticed only a day later in China, and by Korean
astronomers on Oct 13. The supernova, which remained visible for a whole year,
was extensively observed by European astronomers, including Johannes Kepler,
and this SN is often referred to as Kepler's SN. Chinese and Korean astronomers
kept a regular watch on it, and valuable systematic Korean reports over many
months are still preserved, as well as occasional Chinese records. There are no
known Japanese or Arab accounts of this star. The European positional
measurements are far superior to those from East Asia (approximately 1 arcmin
precision as compared with 1 degree). Favourable circumstances assisted the
discovery of the supernova, as it was only about 3 degrees to the north-west of
the planets Mars and Jupiter, which were then in conjunction. This conjunction
was carefully watched by European astronomers in early October of AD 1604, and
was also recorded in China. The peak brightness of the supernova probably did
not occur until late October, so that the supernova was detected well before
maximum.

\begin{figure}
\centerline{\includegraphics[angle=270,width=9cm]{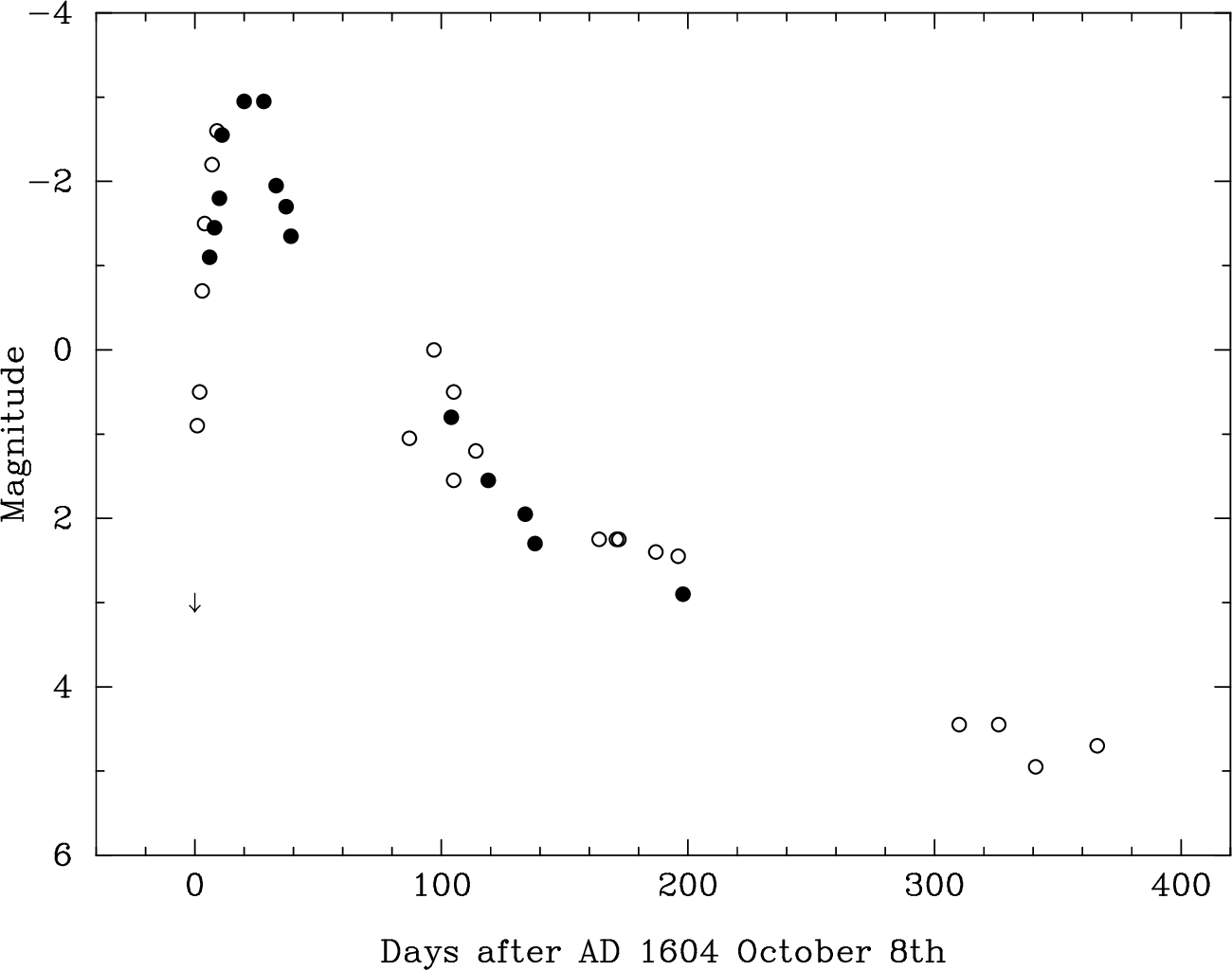}}
\caption{The light curve of SN of AD 1604 from European ($\circ$) and Korean
($\bullet$) observations, with a European upper limit on October
8\label{f:kepler}}
\end{figure}

Chinese observations of the supernova are found in two approximately
contemporary sources: three records in the annalistic {\em Ming Shenzong
Shilu}, and a single record in the dynastic history the {\em Mingshi}. The
guest star was first detected in China on AD 1604 Oct 9 and was finally lost to
sight on AD 1605 Oct 7. Although the guest star was not sighted in Korea until
Oct 13 it attracted considerable attention there. An almost day-to-day record
of the Korean observations of the star over the first six months of visibility
is available, and the regular estimates of brightness parallel the European
observations -- see Fig.~\ref{f:kepler}. Nearly one hundred separate
observations of the guest star are reported in the {\em Sonjo Sillok}. Several
brief accounts of the new star are also to be found in the {\em Chungbo Munhon
Pigo}, a compendium dating from AD 1770.

The most important contemporary European work on the supernova is Johannes
Kepler's {\em De Stella Nova in Pede Serpentarii} which was published in AD
1606 (the more familiar name for {\em Serpentarius} is Ophiuchus). Other
European accounts of the star are to be found in a wide variety of sources (see
\cite{Baa43}). The most important aspects of the European observations are the
accurate position of the star measured by Kepler and its changing brightness
over the twelve months of observation.

We can be fairly confident that AD 1604 Oct 9 was the date of discovery because
several European astronomers, observed the conjunction of Mars and Jupiter on
Oct 8 and did not notice anything remarkable. Due to poor weather, Kepler did
not start observing the supernova until Oct 17. He measured the angular
distance between the new star and several planets and reference stars using a
sextant. Several European astronomers estimated the brightness of the supernova
in the days leading up to maximum. It can probably only be concluded that the
brightness considerably exceeded that of Jupiter, with peak magnitude close to
$-3.0$. Although Kepler's SN has been identified as type~I in the past -- on
the basis of its light curve from the historical observations -- this
classification cannot be justified, as the light curves of some type~I and
type~II SN can be quite similar (see also \cite{Sch96}). With an estimated date
of peak brightness of approximately Oct 28 in AD 1604, the supernova was
detected nearly twenty days before maximum. Using the calculated position for
the new star from Kepler's observations, Baade \cite{Baa43} was able to locate
the remnant as a patch of optical nebulosity with the 100-inch reflector at
Mount Wilson. Subsequently the remnant -- \SNR(4.5)+(6.8) -- has been revealed
as a limb-brightened `shell' supernova remnant at radio and X-ray wavelengths.

%------------------------------------------------------------------------------%
\subsection{Tycho's SN of AD 1572}

The supernova which appeared in the constellation of Cassiopeia during the late
autumn of AD 1572 was compared by observers in both Europe and East Asia with
Venus, and was visible in daylight. Since the most detailed observations of the
supernova in Europe were made by Tycho Brahe, this supernova is often referred
to as Tycho's SN.

Five Chinese records of this supernova are preserved, two in the same sources
as the observations of the supernova of AD 1604: the {\em Ming Shenzong Shilu}
and in the astronomical section of the {\em Mingshi}. Further brief mentions
occur in the biographical section of the {\em Mingshi} and also the {\em Ming
Shigao}, the draft version of the {\em Mingshi}. A fifth account is in the {\em
Yifeng Xianzhi}, a provincial history. Only two brief Korean reports of the AD
1572 star are available. The guest star of AD 1572 was discovered in Korea on
Nov 6 and sighted two days later in China. Noticeably fading by AD 1573 March,
the star finally disappeared from sight some time between Apr 21 and May 19 in
AD 1574. The duration of visibility was thus about 18 months. Chinese records
assert that the star was visible in daylight, while in Korea its brightness was
compared with Venus. The position of the supernova is not defined very
precisely by the East Asian records, but it is notable that its position is
marked on at least two independent Chinese star charts.

The supernova was probably detected in Europe by Maurolycus, abbot of Messina,
on AD 1572 Nov 6 (if not a day or two earlier). It was first observed by Tycho
on Nov 11 (although he remained sceptical of the star's existence until he had
questioned both the servants who were with him and passers by). Tycho
immediately realised this was a new star, and ``began to measure its situation
and distance from the neighbouring stars of Cassiopeia and to note extremely
diligently those things which were visible to the eye concerning its apparent
size, form, colour and other aspects''. Over the following year he was to
make many measurements of the angular distances of the star from the nearby
stars of Cassiopeia, and also determined its distance from Polaris. The
accurate observations by Tycho Brahe and others, which established the fixed
nature of the star, have proved of great importance to modern astronomers in a
different guise: fixing the precise location of the supernova to within a few
arcminutes. Tycho concluded that the new star was situated far beyond the Moon
and among the fixed stars. Hence the supernova contravened the widely accepted
Aristotelian doctrine that change could only occur in the sub-lunar region.
Virtually all the important European observations of the supernova are
contained in Tycho's {\em Astronomiae instaurate progymnasmata} (`Essays on the
new of astronomy'), published in AD 1602. From Korean and European comparisons
of the supernova with Venus, it seems that the peak magnitude of the supernova
was around $-4.0$. Tycho was evidently the only astronomer to carefully watch
the decline in brightness of the new star. In the months after discovery he
successively compared it with Jupiter, then stars of fainter magnitudes.

The remnant of Tycho's SN was first tentatively identified in 1952
\cite{Han52}, when a radio source was found near the then available position
for the SN. This was subsequently confirmed by later radio observations, which
also led to the identification of the faint optical nebulosity associated with
the radio source (Minkowski, private communication in \cite{Bal57}). At radio
and X-ray wavelengths this remnant -- often called either 3C10 or
\SNR(120.1)+(2.1) -- shows a limb-brightened shell $\sim 8$ arcmin in diameter.

\begin{figure}
\centerline{\includegraphics[width=6cm]{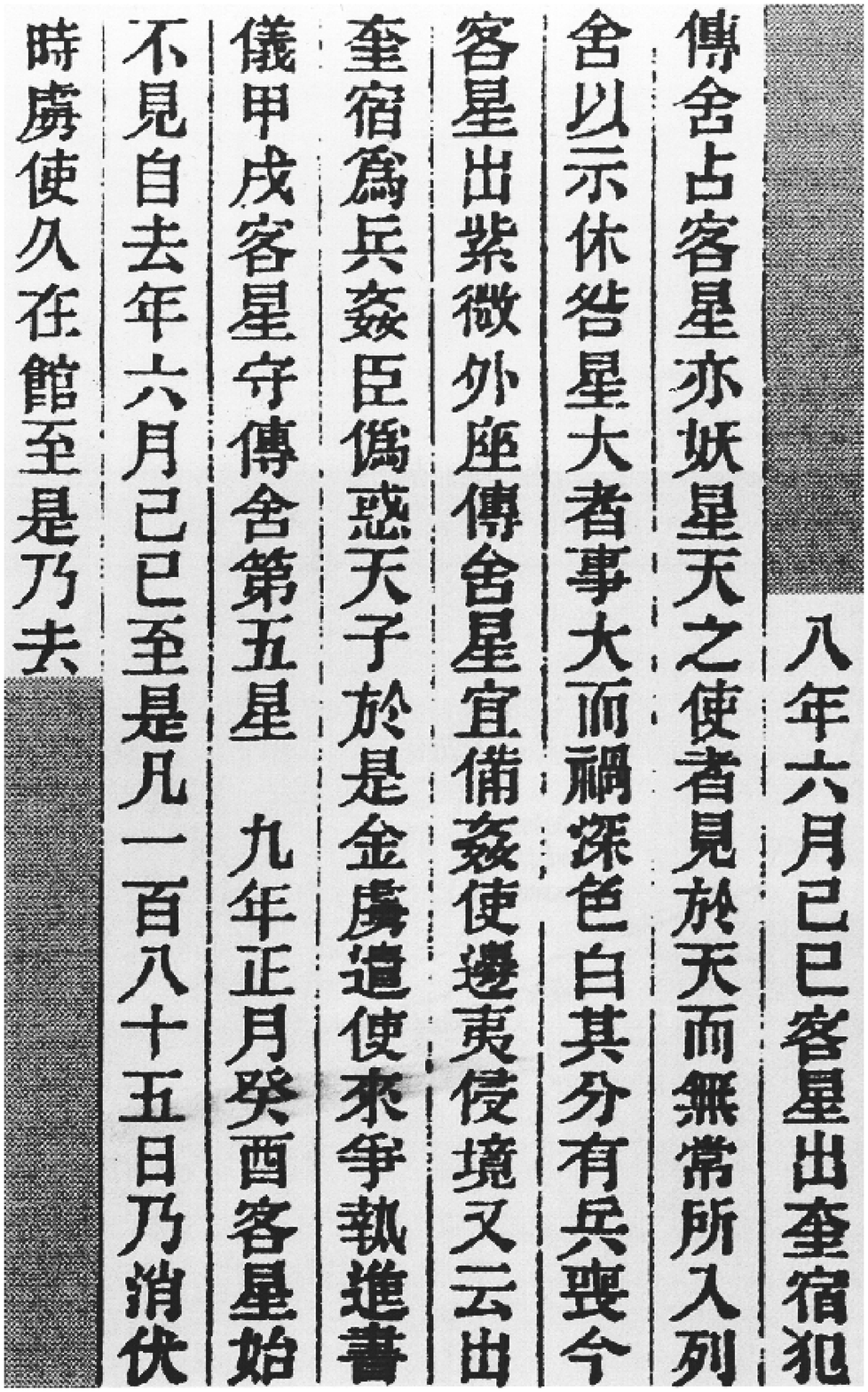}}
\caption{Detailed account of the supernova of AD 1181 from the {\em Wenxian
Tongkao}, which was compiled about a century after the
supernova\label{f:wenxian}}
\end{figure}

%------------------------------------------------------------------------------%
\subsection{The SN of AD 1181}

The new star of AD 1181, which was extensively observed in both China and
Japan, was seen for fully six months. Such a lengthy duration of visibility in
the various historical records is indicative of a supernova. There are three
Chinese records of the new star of AD 1181, from both the North (Jin) and
South (Song) Chinese empires in existence at that time, and five Japanese
accounts. None of these sources report any motion of the star. The most
detailed surviving Chinese account of the guest star is found in the {\em
Wenxian Tongkao} (`Comprehensive study of civilisation'), an extensive work
compiled around AD 1280 -- see Fig.~\ref{f:wenxian}. This account clearly
states that the star was first seen on AD 1181 Aug 6, and was visible for
185 days in total. There is also some valuable positional information in the
record, giving the approximate location of the guest star, and describing it
as guarding the fifth star of the {\em Chuanshe} asterism. The Japanese
records of this guest star come from a variety of sources, including a
retrospective account of the star (and also the new stars of AD 1054 and AD
1006) written in AD 1230. The date of discovery of the guest star in Japan is
one day after its discovery in South China. Other Japanese accounts of the
star occur in various histories, and in diaries of imperial courtiers. Unlike
the Chinese accounts, there are no estimates of the full duration of
visibility of the guest star in the available Japanese accounts, although one
states it was still visible two months after discovery.

The guest star was said to `invade' the {\em Chuanshe} asterism, and to guard
the fifth star of that asterism. The identification of the fifth star with SAO
12076 was proposed by Liu Jinyu \cite{Li83}, in which case the position of the
SN can be deduced as within about 1 degree of SAO 12076; equivalent galactic
coordinates are $l \approx 130^\circ$, $b \approx +3^\circ$. This position is
close to the radio source 3C58 (\SNR(130.7)+(3.1)), which was first proposed as
the remnant of this guest star in 1971 \cite{Ste71}. 3C58 is a `filled-centre'
supernova remnant, in which a pulsar with a period of $\approx 65.58$ ms has
recently been identified (see \cite{Mur01}).

%------------------------------------------------------------------------------%
\subsection{The SN of AD 1054 that produced the Crab Nebula}

The Crab Nebula, which has been known optically since the early 18th century,
was first suggested as the remains of the Chinese guest star of AD 1054 in the
1920s (see \cite{Lun21,Hub28}). A substantial number of records of the guest
star, from both China and Japan, were later assembled by the Dutch sinologue
Duyvendak in 1942 \cite{Duy42}, since when the Crab Nebula has generally been
recognised as the remnant of the oriental guest star of AD 1054. The Crab
Nebula is one of few Galactic SNRs which are known to contain a pulsar -- which
is extremely important for explaining the energetics and structure of the whole
SNR -- and is the prototype of the class of `filled-centre' SNRs. The nebula
was first identified as a source of radio waves in 1963 and in X-rays in 1964,
while the discovery of a pulsar within it in 1968 attracted huge interest
internationally. No other supernova remnant has achieved such notoriety, or
been the subject of so many research papers.

There are many Chinese records of the guest star of AD 1054. It was first
sighted in China at daybreak in the eastern sky on AD 1054 Jul 4 and remained
visible until AD 1056 Apr 6. Three records of the guest star from Japan are
known, apparently from two independent sources. As noted above, one Japanese
record, from AD 1230, also includes discussions of the guest stars of AD 1006
and 1181. Both Chinese and Japanese sources agree that the guest star appeared
close to {\em Tianguan}, which is identified with $\zeta$ Tau. There is no hint
of any motion; on the available evidence, the guest star remained fixed for the
whole of the very long period of visibility. In noting that the star was
visible in daylight (probably for 23 days), Chinese astronomers compared it
with Venus. Japanese observers compared the brightness with Jupiter, again
implying a brilliant object.

It is also likely that this supernova was recorded in Constantinople. Ibn
Bu\d{t}l{\=a}n, a Christian physician, provides a brief record of a new star
seen at this time. Although there have been some suggestions that there are
European records of the supernova of AD 1054, there appears to be no definite
report of it from Europe. It has also been suggested that this supernova, which
was close to the ecliptic, is recorded in cave paintings in the American
south-west which depict a crescent close to a circle or star symbol. However,
only a very approximate date range can be deduced for the paintings (tenth to
twelfth centuries AD), while only one of the pictures shows the correct
orientation of the crescent relative to the new star. If the paintings are
indeed astronomical, which is open to speculation, they might possibly
represent one or more close approaches of the Moon to Venus over the estimated
date range of some two centuries.

%------------------------------------------------------------------------------%
\subsection{The bright SN of AD 1006}

The new star which appeared in AD 1006 was extensively observed in China and
Japan, and was also recorded in Europe and the Arab dominions. The various
records indicate that the star was of extreme brilliance and had an
exceptionally long duration of visibility -- several years.

The reports from China are by far the most detailed, giving not only a fairly
accurate position, but also demonstrating that the new star was certainly seen
for at least three years. Chinese observations are preserved in a wide variety
of sources, including dynastic histories, chronicles and biographies. The new
star was independently sighted in Japan, where it was consistently described as
a {\em kexing} or `guest star' in several independent reports. Discovery in
both China and Japan took place on AD 1006 May 1. Chinese records indicate that
the star remained visible until some time during the lunar month between Aug 27
and Sep 24, when it set heliacally. However, Japanese reports may imply
visibility to Sep 21. After recovery in China on AD 1006 Nov 26, the star was
seen until the following autumn (between Sep 14 and Oct 13 in AD 1007), when it
was lost to view in the evening twilight. The star was evidently sighted again
at dawn some time toward the end of AD 1007, or the beginning of AD 1008, and
-- after a further conjunction with the Sun near the and of AD 1008 -- was
apparently still visible well into the year AD 1009. In China, the extreme
brilliance of the star was emphasised in several ways: it was ``huge$\dots$
like a golden disk'', ``its appearance was like the half Moon and it had
pointed rays''; ``it was so brilliant that one could really see things clearly
(by its light).''  In Japan, the only direct brightness comparison was with the
planet Mars, although the fact that the new star made such a profound
impression on the imperial court suggests that it was a remarkable sight.

Brief Arab reports of the new star are preserved in chronicles from several
regions: Egypt, Iraq, north-west Africa or Spain, and Yemen. The most likely
date for the discovery of the new star in the Arab dominions is AD 1006 Apr 30,
one day earlier than in China and Japan. Furthermore several Arab records
suggest that the star disappeared around Sep 1, a few weeks before it ceased to
be reported in Japan. Two accounts from Europe -- in the chronicles of the
monasteries at St Gallen in Switzerland and at Benevento in Italy -- clearly
refer to the new star, the former source indicating that it was visible for
three months. Several other annals note the occurrence of a `comet' in or
around AD 1006. Since there is no notice of a comet in Chinese history around
this time, it may be {\em presumed} that the European chroniclers also referred
to the new star but had no separate term to describe a brilliant star-like
object. The St Gallen record mentions frequent interruptions in visibility,
which imply that the star no more than skimmed the southern horizon; this
provides a valuable declination limit for the position of the supernova.

The identification of the likely remnant of this SN was made in 1965
\cite{Gar65}, when a search was made of radio catalogues covering part of the
region for the SN from the historical observations. The radio source PKS
1459$-$51 -- also known as MSH 14$-$4{\em 15}, or \SNR(327.4)+(14.6) from its
Galactic coordinates. Subsequent improved observations confirmed this as the
remnant, when its structure was revealed to be a limb-brightened `shell'
supernova remnant about half a degree in diameter.

%------------------------------------------------------------------------------%
\section{Probable historical Supernovae before AD 1000}\label{s:older}

Other long duration guest stars recorded before AD 1000 in China that are
possibly records of supernovae are from AD 393, 386, 369 and 185, which are
briefly discussed here.

\begin{figure}
\centerline{\includegraphics[width=8cm]{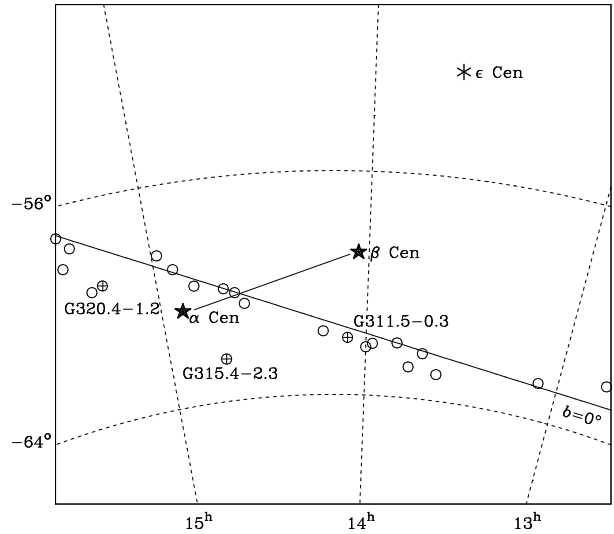}}
\caption{The region of the {\em Nanmen} asterism, near which the SN of AD 185
occurred. The positions of the bright stars $\alpha$ and $\beta$ Cen which are
thought to comprise {\em Nanmen} are shown, along with the fainter $\epsilon$
Cen. The circles indicate the centroids of catalogued Galactic SNRs, some with
additional crosses and labels. The solid line shows the Galactic
equator\label{f:nanmen}}
\end{figure}

The new stars of AD 393, 386 and 369 appeared towards the end of the Jin
dynasty in China. All three objects are recorded in the astronomical treatises
of histories of both the Song and Jin dynasties (the {\em Songshu} and {\em
Jinshu}). However, the two records of each of these stars clearly show a common
origin.
%
% AD 393
%
The guest star of AD 393 was visible for some 8 months, and so is likely to
have been a supernova. The position of this star is within the {\em Wei}
asterism, which lies near the Galactic equator. There are several SNRs within
this region, and it is not possible to identify the remnant of this SN
unambiguously.
%
% AD 386
%
The new star of AD 386 was visible for anything from about 60 to 115 days.
Since the duration of visibility may have been rather short, the possibility
that this was a nova cannot be ruled out. Nevertheless, a supernova
interpretation is also plausible. The position of the guest star is not well
defined -- it could have appeared near the star group {\em Nandou} (which lay
close to the Galactic equator), or was merely somewhere in the range of RA
defined by {\em Nandou}. In the latter case, although the RA would be fairly
well defined, the declination would not. In the former case there are several
possible identifications for the remnant of this guest star, with
\SNR(11.2)-(0.3) perhaps being the prime candidate.
%
% AD 369
%
The guest star of AD 369 is described with only limited details. It was
reported to be visible for 5 months, but only poorly localised position is
reported. If the star was near the galactic equator a supernova interpretation
would be plausible, but if it were far from the galactic plane it is more
likely a slow nova.

The earliest new star which merits investigation as a possible supernova was
seen in China in AD 185. This event is reported only in a single early source,
the {\em Hou Hanshu}, which was composed towards the end of the third century
AD. The new star was recorded as being visible for at least 8 months, or
possibly even 20 months (depending on interpretation of part of the record to
mean `next year' or `year after next'). The star was reported to be within the
{\em Nanmen} asterism. Although some authors have questioned the identification
of this asterism, comparison with contemporary records and star charts supports
the usual identification of $\alpha$ and $\beta$ Cen with {\em Nanmen}, which
lies close to the Galactic equator. \SNR(315.4)-(2.3) is the prime candidate
for the remnant of the SN of AD 185 among the SNRs in the present catalogue of
Galactic SNRs between $\alpha$ and $\beta$ Cen -- see Fig.~\ref{f:nanmen} --
although it should be noted
that there are other remnants in this region which have not yet been studied in
great detail.

%------------------------------------------------------------------------------%
\section{Other possible and spurious Supernovae}\label{s:not}

\subsection{Did Flamsteed see the Cas~A supernova in AD 1680?}

\begin{figure}
\centerline{\includegraphics[width=7cm]{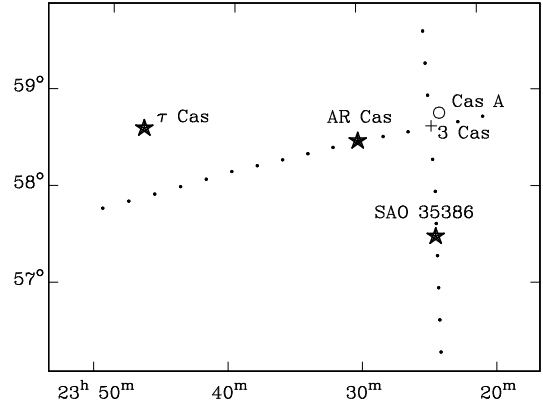}}
\caption{Plot of the relative positions of Cas~A, AR Cas and SAO 35386 (epoch
J2000.0).  Also shown are $\tau$ Cas and the site of `3~Cas', and dotted lines
to indicate the arcs measured by Flamsteed from $\beta$ Peg and $\beta$
Per\label{f:3cas}}
\end{figure}

Cassiopeia~A (Cas~A) is an obviously young and relatively nearby SNR, which is
a bright source at radio and X-ray wavelengths, showing a clumpy shell of
emission. Optically Cas~A shows a patchy ring of many expanding knots. The date
at which the knots would converge, assuming no deceleration, is AD 1671 (see
\cite{Tho01} for a recent review). These observations are consistent with Cas~A
being produced by a SN in AD 1671 or shortly afterwards, provided these optical
knots have undergone only a very small deceleration subsequently, which is
expected if these optical knots are very dense ejecta from the SN. The distance
to Cas~A can be found trigonometrically to be $3.4^{+0.3}_{-0.1}$ kpc, by
combining the proper motion of the optical filaments in the plane of the sky
with their radial velocities measured spectroscopically. Given the likely date
and proximity of the supernova that produced the Cas~A SNR, it has been a
puzzle that no historical observations of it are available. In 1980 Ashworth
suggested \cite{Ash80} that the supernova that produced Cas~A was indeed
recorded by Flamsteed in AD 1680, as he catalogued a 6th magnitude star
`3~Cas', to the west of $\tau$ Cas, fairly close to the present site of Cas~A,
where there is no known star today. However, the discrepancy in the positions
of 3~Cas and Cas~A  -- about 10~arcmin -- is very much larger than Flamsteed's
typical measurement error. Alternatively -- as proposed by Broughton
\cite{Bro79} and by Kamper \cite{Kam80} -- Flamsteed did not observe the
supernova, but instead accidentally compounded his measurements of two separate
stars (AR Cas and SAO 35386), neither of which he actually catalogued. Since it
is possible to identify the other faint stars that would have produced the
erroneous 3~Cas position -- see Fig.~\ref{f:3cas} -- with measurement
uncertainties that are plausible for Flamsteed's typical precision, it seems
most unlikely that Flamsteed observed the supernova which produced Cas~A.

\subsection{The Korean guest stars of AD 1592}

During a period of about one month in AD 1592, four separate guest stars were
reported in Korea in the {\em Sonjo Sillok}, the official annals of the King
Sonjo who reigned in Korea from AD 1567 to 1608. The first of these appeared in
Cetus and was observed for 15 months. Two more guest stars (both in Cassiopeia)
were seen for periods of at least three and four months, while a fourth (in
Andromeda) was visible for more than a month. In each case the position
remained unchanged; small refinements in the recorded locations of the three
stars of longest duration only serve to emphasise their fixed nature.
Surprisingly, none of these objects was reported in China or Europe, which
suggests that all were by no means brilliant: probably of 2nd magnitude or
fainter. The various records have been investigated in detail \cite{Ste87}. In
summary, it would appear that as many as four novae may have occurred in AD
1592, but in no instance is a supernova interpretation tenable.

\subsection{The spurious supernovae of AD 1408, 1230 and 837}

In 1979 Li Qibin \cite{Li79} assembled several Chinese records of two temporary
stars observed in AD 1408. Six of these accounts were from Szechuan province
and described a bright star which appeared in the east, most probably on Sep
10. Three further reports were in official histories of China and related to a
star which appeared on Oct 24 and `did not move'. Li Qibin regarded the two
objects as identical and proposed a supernova identification. Subsequently
Imaeda \& Kiang \cite{Ima80} found two further Japanese records mentioning a
guest star on a date equivalent to AD 1408 Jul 14. Although no position was
recorded for this object, it was inferred that the observation represented an
earlier sighting of the stars seen in China on Sep 10 and Oct 24. They further
concluded that the star ``was quite likely to be a supernova explosion''. The
publications by Li Qibin and Imaeda \& Kiang led to consideration of
\SNR(69.0)+(2.7) (CTB~80) as the remnant of the star by various authors.
However, it has been shown \cite{Ste86} that the `star' of Sep 10 was merely a
meteor, and also there are insufficient grounds for linking the guest star of
Jul 14 seen in Japan with the star appearing on Oct 24 as reported in China.

Wang Zhenru in 1987 suggested \cite{Wan87} that a `bushy star' seen for more
than three months in AD 1230 was a supernova, and further proposed a
$\gamma$-ray source 2CG 054$+$01 as its remnant. (Wang Zhenru also proposed an
association between a purported 14th century BC supernova record with another
$\gamma$-ray source, but this is highly speculative.) The object was, in fact,
a comet. Ho Peng Yoke \cite{Ho62} had already drawn attention to the motion of
the same object as described in the astronomical treatise of the {\em Jinshu}
-- which contained records from North China.

Two guest stars appeared in AD 837, which were discovered soon after Halley's
Comet had been detected in that year. Various authors have interpreted the
records of one of these guest stars as evidence of a supernova, which they
associate with the SNR \SNR(189.1)+(3.0) (IC443). Although the first star was
fairly close to the galactic equator, the duration of visibility (22 days) was
very short. Further, the star disappeared while still some 7 hours in RA to the
east of the Sun, so that its visibility would not be impaired by the twilight
glow. A supernova interpretation can thus be rejected; the star was most likely
a fast nova. The second star, visible for 75 days in high galactic latitude was
evidently also a nova.

%------------------------------------------------------------------------------%
\section{Future Prospects}\label{s:future}

Looking to the future, it seems unlikely that records of additional supernovae
-- other than those discussed above -- will come to light. Most of the
accessible historical sources, especially those of East Asia, have been fairly
throughly examined. Many medieval Arab and European chronicles are still
unpublished, but even to access a small proportion of this material, which is
scattered in numerous archives, would be extremely time-consuming. Furthermore,
chroniclers were mainly interested in reporting only the most spectacular
events. Hence although it would seem likely that further accounts of the
brilliant supernova of AD 1006 might well emerge, the prospects for fainter
objects -- including the supernova of AD 1054 -- would appear to be far from
promising. In particular, caution should be exercised in assessing the
viability of further potential records of historical supernovae.

The remnants of the supernovae observed since AD 1000 are well-established.
However, improved distance measurements for the suggested remnants of the
proposed supernovae of AD 393, 386 and 185 would be valuable. These results
would lead to better estimates of the physical size and hence age, which might
help distinguish between individual candidate remnants.

%==============================================================================%

%==============================================================================%

\end{document}